\documentclass[10pt]{article}
\usepackage{geometry}
\usepackage{setspace}
\usepackage[numbers]{natbib}
\usepackage{doc}
\usepackage{url}
\usepackage{graphicx}
\usepackage{epstopdf}
\usepackage{hyperref}
\usepackage{amsthm}
\usepackage[font=small,labelfont=bf]{caption}
\title{Approximation Algorithms for the Maximum Profit Pick-up Problem with Time Windows and Capacity Constraint\footnote{This research was partially supported by NSF award IIP1439718 and CPRIT award RP150164}}
\author{Bogdan Armaselu$^1$ and Ovidiu Daescu$^1$}
\date{}

\begin{document}

\maketitle
\begin{center}
$^1$Department of Computer Science, The University of Texas at Dallas, Richardson, TX USA

\texttt{\{bxa120530, daescu\}@utdallas.edu}
\end{center}

\abstract{
In this paper, we study the Maximum Profit Pick-up Problem with Time Windows and Capacity Constraint (MP-PPTWC). 
Our main results are 3 polynomial time algorithms, all having constant approximation factors. 
The first algorithm has an approximation ratio of $\simeq 46 (1 + (71/60 + \frac{\alpha}{\sqrt{10+p}}) \epsilon) \log T$, where: 
(i) $\epsilon > 0$ and $T$ are constants;
(ii) The maximum quantity supplied is $q_{max} = O(n^p) q_{min}$, for some $p > 0$, where $q_{min}$ is the minimum quantity supplied;
(iii) $\alpha > 0$ is a constant such that the optimal number of vehicles is always at least $\sqrt{10 + p} / \alpha$. 
The second algorithm has an approximation ratio of $\simeq 46 (1 + \epsilon + \frac{(2 + \alpha) \epsilon}{\sqrt{10 + p}}) \log T$. 
Finally, the third algorithm has an approximation ratio of $\simeq 11 (1 + 2 \epsilon) \log T$. 
While our algorithms may seem to have quite high approximation ratios, in practice they work well and, in the majority of cases, the profit obtained is at least 1/2 of the optimum.
 }

~

\begin{centering}
\end{centering}

\section{Introduction}
\label{introduction}

The Maximum Profit Pick-up Problem with Time Windows and Capacity Constraint (MP-PPTWC) is stated as follows. 
We are given a certain product with a unit price of 1, and set of $n$ suppliers (or sites), each site $i$ being specified by its coordinates $(x_i, y_i)$ in the plane and an interval $[e_i, l_i]$ called "time window". 
A vehicle can only visit a site $i$ during its time window $[e_i, l_i]$. If a vehicle reaches a site $i$ at a time $t < e_i$, it has to wait until time $t = e_i$. 
The values $e_i, l_i$ are assumed to be integers in the interval $[0, T]$, where $T$ is a given constant. 
Each site $i$ has a constant quantity $q_i$ of the product available.
We are also given a depot $D$ and a vehicle type with a capacity of $Q$, and unit fuel consumption. The unit price of fuel is also assumed to be 1. 
The distance $d_{ij}$ (in km) between suppliers $i$ and $j$ is given by some metric. The network consisting of all sites and the depot is considered a complete graph.
On every edge of the graph, all vehicles are assumed to travel at the same speed $s_v$. This makes sense since the point-to-point trips, in real world, are composed of mutiple road segments, and have roughly the same average road conditions overall.
The time needed to load the supplies is assumed to be $0$. 
Let $c_{ij} = d_{ij}$ denote the traveling cost from site $i$ to site $j$, and $t_{ij} = d_{ij} / s_v$ denote the traveling time from $i$ to $j$. 
The number of vehicles is unlimited. Each vehicle $k$ is to be given a route $r_k$, starting and ending at $D$, so that it can collect a quantity of $\min \{ Q, \sum_{i \in r_k} q_i \}$.
The routes $r_k$ must be vertex disjoint (except for the depot $D$). That is, only one vehicle is allowed to cisit a site.
The goal is to find $m$ and a set of $m$ routes $r_1, \dots r_m$ for $m$ vehicles, such that the total profit $P$ is maximized, where $P = \sum_{k = 1}^{m} (\sum_{i \in r_k} q_i - \sum_{(i, j) \in r_k} c_{ij})$.

Note that this \textit{pick-up} problem we study is different from the well-known \textit{delivery} problem, where vehicles are required to deliver certain quantities to certain stations and the goal is to minimize the costs. 
To the best of our knowledge, there is no previous work done on the pick-up problem with the goal to maximize the profit, studied in this paper.

The obvious application of this problem is decision making for profit maximization of a certain industry. However, it can be used in many other fields, such as public transportation. 
Stops and depots are fixed, all routes have unit ticket price, and the goal is to assign routes to maximize the total profit (total revenue minus total fuel costs).

The problem is an extension of the Traveling Salesman Problem (TSP), which is known to be strongly NP-hard. Hence, MP-PPTWC is also a strongly NP-hard problem. 
Moreover, it is shown in \cite{Savalsbergh} that finding a feasible solution for a vehicle routing problem with time windows is NP-hard in the strong sense, even for only one vehicle.

\subsection{Related work}
\label{Related work}

Even though any generalization of TSP is NP-hard, approximation schemes have been found for most of them.
Currently, the best approximation algorithm known for the general metric TSP problem is the one given by Christofides \cite{Christofides}, which has an approximation ratio of $3/2$.
In 2000, Arora et. al presented an $O(n (\log n)^{O(c)})$-time, $(1 + 1/c)$ approximation ratio algorithm for the planar version of the Euclidean TSP \cite{Arora}, using a randomized algorithm. 
They also show how to extend their algorithm for the $d$-dimensional case, in which case their running time increases to $O(n (\log n)^{O((\sqrt{d} c)^{d-1})})$.

In 1994, Fisher studied the VRP problem \cite{Fisher94}, in which there are no time windows, all of the $m$ vehicles have an equal capacity $Q$, there is only one depot for all vehicles, and each customer $i$ is specified by its coordinates in the plane and the demand $q_i$ of the product. 
They gave a near-optimal iterative algorithm for the problem, using an iterative lagrangian relaxation of the constraints. In each iteration, a minimum K-tree approach is used to obtain a relaxation, in polynomial time. 

In 1995, Fisher et al \cite{Fisher97} studied the VRPTW problem, which is the same as VRP, but with time windows $[e_i, l_i]$ and different capacity constraints $Q_j$ for the vehicles. 
They show how to adapt their algorithm in \cite{Fisher94} to work for this problem, and they also give another approach using linear programming. 

Perhaps the most relevant problem related to the one that we study is the Multi-Depot Vehicle Routing Problem with Time Windows and Multi-type Vehicle Number Limits \cite{Shang}. 
The main difference is that the goal is to minimize the number of vehicles used (if the problem is feasible) or maximize the number of customers visited on time (if the problem is not feasible). 
The problem was solved by Wang et. al in 2008 using a genetic algorithm approach \cite{Shang}. However, their algorithm is iterative, and has no approximation ratio guaranteed. 

Pick-up Routing Problems have also been studied. Phan et. al \cite{Phan} consider the Pick-up and Delivery Problem with Time Windows and Demands. 
Their optimum criteria are: minimizing total traveled distance, minimizing number of vehicles used and maximizing revenues. 
The difference in our case is that our optimum criterion is maximizing the profit (i.e. revenue minus cost), rather than just revenue. 
Jih et. al \cite{Hsu} study the Pickup and Delivery Problem with Time Window Constraints. The goal of the problem the problem is to minimize both the total traveling time and the total waiting time. 
Both \cite{Phan} and \cite{Hsu} give genetic algorithm approaches.
Exact algorithms for Pick-up Routing Problems, that run in exponential time, have been given by Dessouki et. al \cite{Dessouki}.

In the Prize-Collecting TSP, the goal is to minimize the traveling costs while maximizing the payout. 
Balas et al \cite{Balas} combine these criteria into a single objective, namely to minimize the expenses (i.e. traveling costs minus payout), while obtaining at least a given quota in reward.
The differences from our problem are that: we do not have a reward quota to meet, there are multiple vehicles, capacity constraint, and the problem is to pick up the goods, not to deliver.

Recently, an approximation algorithm for the Deadline-TSP problem was given by Bansal et. al \cite{Bansal}. 
In the Deadline-TSP problem, we are given $n$ sites, where each site $i$ has a deadline $D_i$, and we want to find a tour that maximizes the number of sites that can be visited. The algorithm that they give has an approximation ratio of $O(\log n)$. 
They also show how to extend it to the more general Vehicle Routing with Time Windows problem, in which there are $m$ vehicles and sites also have release times. For this problem, they give an algorithm with an approximation ratio of $O(\log^2 n)$. 
Finally, they give an $O(\log(1/\epsilon))$-approximate algorithm for the $\epsilon$-relaxed version of the problem, in which we are allowed to extend the deadlines by a factor of $1 + \epsilon$.

Several other variations of vehicle routing with time windows have been studied, including Vehicle Routing with unlimited number of vehicles \cite{Desrochers, Gendreau}. 

\subsection{Our contributions}
\label{Our contributions}

We first list the assumptions that need to hold in order for our approximation algorithms to hold.

\textbf{Assumption 1}. If $q_{min}, q_{max}$ are the minimum (resp. maximum) non-zero rewards, then $q_{max} \leq Q$ and $q_{max} = O(n^p) q_{min}$, for some $p > 0$. 

\textbf{Assumption 2}. $\forall i, j, c_{ij} \leq \frac{1}{2} \epsilon q_j$, for some $\epsilon > 0$.

\textbf{Assumption 3}. The optimal number of vehicles is always $m^* \geq \sqrt{10 + p} / \alpha$, for some $\alpha > 0$.


We give 3 algorithms for MP-PPTWC. 

First, we give an $O(n^{10 + p})$ time algorithm that has an approximation ratio of $16 \ln 2 \cdot (1 + \pi) (1 + (71/60 + \frac{\alpha}{\sqrt{10+p}}) \epsilon) \log T \simeq 46 (1 + (71/60 + \frac{\alpha}{\sqrt{10+p}}) \epsilon) \log T$, where $T$ is the latest time of any time window.
The algorithm relies on a novel APX for bin packing, as well as a novel Time Window-TSP approach. 

The second algorithm also runs in time $O(n^{10 + p})$, but uses an APTAS for bin packing and has an approximation ratio of $46 (1 + \epsilon + \frac{(2 + \alpha) \epsilon}{\sqrt{10 + p}}) \log T$. 

Finally, the third algorithm uses well-separated pair decomposition (rather than bin-packing) to split the set of sites into a sequence of pairs of subsets. 
Each vehicle is assigned one of these subsets, then a routing is computed for each vehicle. 
The running time is $O(s^2 n^{9 + p})$ and the approximation ratio is $11 (1 + \frac{\pi}{1 + s}) (1 + 2 \epsilon (1 + 1/(1+s))) \log T$, where $s > 0$ is a constant. 
For $s \geq 5$, this is better than the second algorithm, regardless of the values of $\epsilon$ and $p$. 
As $n \rightarrow \infty$, the approximation ratio is $\simeq 11 (1 + 2 \epsilon) \log T$ (and the running time is $O(\min{n^{10 + p}, n^{9 + 2 p}}))$.

We leave as open problems proving approximation bounds for a few, more general versions of the problem. 

\bigskip

\section{Algorithms for MP-PPTWC}
\label{MP-PPTWC}

We describe our three algorithms for MP-PPTWC.

\subsection{Algorithm 1}
\label{Algorithm 1}

Let $q_i$ be the fixed quantity supplied by site $i$. We first decompose the set of sites into subsets, using a Bin Packing approximation algorithm, with $q_1, \dots, q_n$ as input item sizes, and $Q$ as input bin volume. 
Suppose the algorithm divides the set $\{1, \dots, n\}$ into a sequence of $m$ subsets $S_k = \{ S_{k1}, \dots, S_{ks_k}\}, k = 1 \dots m$. 
For Bin Packing, we use the modified First-Fit Decreasing strategy in \cite{delaVega}. 
We consider $m$ vehicles, and for each vehicle $k$, we assign a subset $S_k$ of sites to be visited, then do the following:

\textbf{1}. Find a route $R_k$ that maximizes the total payout $P_k = \sum_{i \in R_k - \{ D \}} q_i$, using the bi-criteria algorithm in \cite{Bansal}.

\textbf{2}. For every $j$ such that $t_j^k \notin [e_j, l_j]$, we remove $j$ from the route. If $j', j''$ are the predecessor, respectively, the successor of $j$ in $R_k$, we add an edge $j' -> j''$ to $R_k$.

We now analyze the performance of our algorithm. Let $m^*$ be the optimal number of vehicles needed. 
For every vehicle $k$, let $d_k^*$ denote the optimal length of any tour that visits all sites in $S_k$ starting and ending at $D$, and $d^*$ the optimal total length of any set of $m^*$ such tours (over all possible $S_k$'s).

The first two lemmas give a bound on the ratio between the total distance traveled by the routing output by the algorithm, and the optimal total distance.

~

\textbf{Lemma 1}. $\sum_{k=1}^m d_k^* \leq \frac{m}{m^*} (1 + \pi) d^*$.

\textbf{Proof}. 
Consider each subset $S_k$. 
First suppose all sites lie on a circle of radius $r$ centered at the depot, and sites define a regular polygon of length $l$. 
It is easy to see that 
\begin{equation}
\sum_k d_k^* \leq m ((n - m) l + 2 r),
\end{equation}
since the optimal tour for any $S_k$ is the regular polygon formed by the sites, plus another $2 r$ to connect the depot. 
On the other hand, one can see that 
\begin{equation}
d^* \geq 2 m^* r + (n - m^*) l,
\end{equation}
as the optimal $m^*$ tours are formed by disjoint chains of the regular polygon, plus $2 r m^*$ to connect the depot in each tour. See figure 1 for an illustration. 
We therefore get, in this case, 
\begin{equation}
\frac{\sum_{k=1}^m d_k^*}{d^*} \leq \frac{m ((n - m) l + 2 r)}{2 m^* r + (n - m^*) l} = \frac{2 m r + m l (n - m)}{2 m^* r + l (n - m^*)}
\end{equation}
From $l = 2 r \sin(\pi/n) \leq 2 r \pi / n$, we get 
\begin{equation}
\frac{\sum_{k=1}^m d_k^*}{d^*} \leq \frac{2 m r (1 + \pi - \pi m / n)}{2 m^* r (1 - \pi/n + \pi/m^*)} \leq \frac{m (1 + \pi)}{m^*}.
\end{equation}
Now suppose we add a site $i$ that is not on the circle. We want to prove that the approximation ratio is even less in this case. 
Indeed, $\sum_{k=1}^m d_k$ increases by $a + b - m l$, where $a, b$ are the distances to the closest two sites in some $S_k$. 
On the other hand, $d^*$ increases by $a_1 + b_1 - l$, where $a_1, b_1$ are the distances to the closest two sites on the circle. 
It is easy to check that 
\begin{equation}
a + b - m l \leq (a_1 + (m - 1) / 2 l) + (b_1 + (m - 1) / 2 l) - m l = a_1 + b_1 - l.
\end{equation}
Since $i$ is arbitrarily chosen, it follows that in any other configuration of the sites the approximation ratio is at most as large. 
That is, 
\begin{equation}
\sum_{k=1}^m d_k^* \leq \frac{m}{m^*}(1 + \pi) d^*
\end{equation} holds in any possible scenario. \qed

~

\textbf{Lemma 2}. $\sum_{k=1}^m d_k^* \leq (71/60 + \frac{\alpha}{\sqrt{10+p}}) (1 + \pi) d^*$.

\textbf{Proof}. 
The First-Fit-Decreasing approach in \cite{delaVega} yields a number of bins $m \leq 71/60 m^* + 1$, where $m^*$ is the optimal number of bins. 
By Assumption 3, we get 
\begin{equation}
m \leq m^*(71/60 + \frac{\alpha}{\sqrt{10+p}}).
\end{equation}
Plugging $\frac{m}{m^*}$ into the equation in Lemma 1, we get 
\begin{equation}
\sum_{k=1}^m d_k^* \leq (71/60 + \frac{\alpha}{\sqrt{10+p}}) (1 + \pi) d^*. 
\end{equation} \qed

~

\begin{figure}[t]
	\centering
	\includegraphics[scale=0.8]{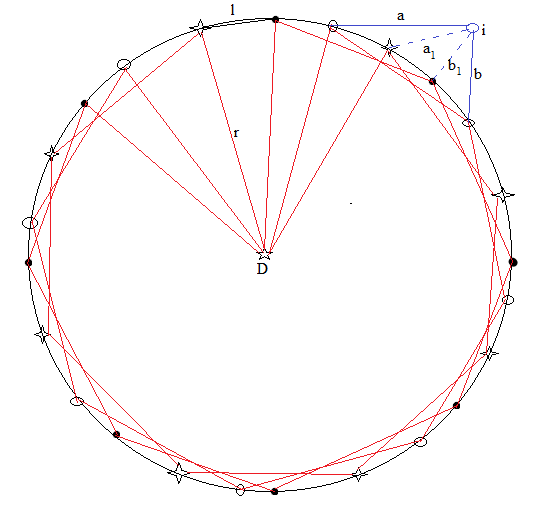}
	\caption{All sites are vertices of a regular polygon centered at the depot. Sites are divided into 3 subsets. 
		The optimal routing for the given division of the sites is displayed. If we add a site $i$, the length of the routing increases by $a + b - m l$.}
\end{figure}

The following lemma gives an upper bound on the ratio between the optimal profit obtainable by a route on a given set $S_k$ and the actual profit obtained by our algorithm.

\textbf{Lemma 3}. 
For every vehicle $k$, Algorithm 1 computes a route $R_k$ with a revenue $P_k \geq \frac{P_k^*}{8 (\ln 2) \log T (1 + \epsilon)}$, where $P_k^*$ is the revenue obtained by an optimal route within $S_k$. 
The running time is $O(n^{10})$.

\textbf{Proof}. 
In step 1, the bi-criteria approximation algorithm in \cite{Bansal} obtains a revenue of at least
\begin{equation}
\sigma_k \geq \frac{\sigma_k^*}{8 \ln 2 \log T}, 
\end{equation} where $\sigma_k^*$ is the optimal revenue obtainable within $S_k$. 
From Assumption 2, it follows that 
\begin{equation}
c(k) \leq 1/2 \epsilon \sum_{i \in R_k} q_i \leq 1/2 \epsilon \sigma_k, 
\end{equation} so the profit cannot increase if we remove sites with non-zero reward and replace them with other sites, such that the total distance is reduced. 
Hence, the profit of our computed $R_k$ is 
\begin{equation}
P_k = \sigma_k - c(k) \geq \frac{\sigma_k^*}{8 \ln 2 \log T} \cdot (1 - \epsilon / 2) \geq \frac{\sigma_k^*}{8 \ln 2 \log T (1 + \epsilon)} \geq \frac{P_k^*}{8 \ln 2 \log T (1 + \epsilon)}. 
\end{equation}
The running time of the algorithm in \cite{Bansal}, for each $k$, is $O(N_k^{9} T \log T)$, where $N_k$ is the number of vertices of $G_k$. 
Since $T$ is a constant and $N_k = O(n)$, this yields $O(n^{10})$ in our case. All other running times are dominated by this one. \qed

~

\textbf{Lemma 4}. 
Let $\rho_k$ be the reward obtainable by an optimal routing on $S_k, \forall k$, and $\rho^*$ be the reward obtainable by an optimal routing. 
Then $\sum_{k=1}^{m} N_k \geq \frac{N^*}{2}$.

\textbf{Proof}. 
Let $R' = \cup_{k = 1}^{m'} R_k'$ be a routing that maximizes the total reward, and $R_k$ be a routing that maximizes the reward obtainable from $S_k, \forall k = 1 \dots m$. 
It is known \cite{Vazirani} that any Bin-Packing algorithm will use at most $2 m^*$ bins, where $m^*$ is the optimal number of bins. Therefore, $m' \leq 2 m$. 
From $R'$, we remove the sites from the $m' - m$ routes with fewest sites to visit, and re-insert them into the $m$ routes with the most customers. If capacity constraints are violated, we leave out the least rewarding sites.
For each such site removal and re-insertion, the number of sites that cannot be reached within their time window increases by at most one. For each vehicle, we construct the route such that the least rewarding site is left out.
Since we move at most $n/2$ sites in this way, we get a reward of 
\begin{equation}
\sum_{k=1}^{m} \rho_k' \geq \frac{\rho^*}{2}. 
\end{equation}
Since $R_k$ is optimal for vehicle $k$, we get 
\begin{equation}
\sum_{k=1}^{m} \rho_k \geq \sum_{k=1}^{m} \rho_k' \geq \frac{\rho^*}{2}.
\end{equation}  \qed

~

Using Lemmas 1-4, we can prove the following result.

\textbf{Theorem 5}. 
MP-PPTWC can be solved within an approximation ratio of $16 \ln 2 \cdot (1 + (71/60 + \frac{\alpha}{\sqrt{10}}) \epsilon) (1 + \pi) \log T \simeq 46 (1 + (71/60 + \frac{\alpha}{\sqrt{10+p}}) \epsilon) \log T$ in $O(n^{10})$ time using Algorithm 1.

\textbf{Proof}. 
Because of our Bin-Packing algorithm, we never use more than $(71/60 + \frac{\alpha}{\sqrt{10+p}}) m^*$ vehicles, where $m^*$ is the optimal number of vehicles.
Since (by Lemma 2) the travel distance is at most $1 + \pi$ of the optimal (for the same number of vehicles), we obtain a reward of $\frac{1}{1 + \pi}$ of the optimal due to traveling distance approximation. 
By Lemma 4, for each vehicle, if we used optimal reward-maximization algorithm, we would obtain at least $1/2$ of the optimal reward, due to the bin packing of sites to specific vehicles. 
By Lemma 3, we obtain at least $\frac{1}{8 \ln 2 \log T (1 + \epsilon)}$ of this amount using the approximate reward-maximization algorithm. 
Putting it all together, we obtain at least 
\begin{equation}
\frac{1}{1 + \pi} \cdot \frac{\sum_{k=1}^{m} P_k^*}{(1 + (71/60 + \frac{\alpha}{\sqrt{10+p}}) \epsilon) 8 \ln 2 \log T} \geq \frac{1}{2(1 + \pi)} \cdot \frac{P^*}{8 \ln 2 (1 + (71/60 + \frac{\alpha}{\sqrt{10+p}}) \epsilon) \log T}
\end{equation}
as profit, where $P^*$ is the optimal profit. 
Hence, the approximation ratio is 
\begin{equation}
16 \ln 2 (1 + \pi) (1 + (71/60 + \frac{\alpha}{\sqrt{10+p}}) \epsilon) \log T.
\end{equation}
The running time is dominated by the time needed to compute the approximate routes $R_k$, which, by Lemma 3, is $O(n^{10 + p})$. \qed

\subsection{Algorithm 2}
\label{Algorithm 2}

Algorithm 2 is the same as Algorithm 1, except it uses a different Bin Packing approach for partitioning the set of sites into subsets. Namely, it uses the $O(n^{4/\eta^2})$ time Asymptotic PTAS algorithm in \cite{delaVega} that yields a number of bins not exceeding $(1 + \eta)m^* + 1$, where $m^*$ is the optimal number of bins.

We now analyze the performance of this algorithm. In this sense, we reuse the notations in Lemmas 1-4, as well as Theorem 5.

~

\textbf{Lemma 6}. The Bin Packing step of Algorithm 2 yields a number of vehicles not exceeding $(1 + \frac{2 + \alpha}{\sqrt{10 + p}})m^*$ in $O(n^{10 + p})$ time. 

\textbf{Proof}. 
The APTAS Bin Packing algorithm yields a number of bins $m \leq (1 + \eta)m^* + 1$. Let $\eta = \frac{2}{\sqrt{10 + p}}$. 
Thus, \begin{equation}
m \leq (1 + \frac{2}{\sqrt{10 + p}}) m^* + 1.
\end{equation}
By Assumption 3, we get 
\begin{equation}
m \leq (1 + \frac{2 + \alpha}{\sqrt{10 + p}}) m^*. 
\end{equation}
The running time is $O(n^{4/\eta^2}) = O(n^{10 + p})$. \qed

~

The following lemma is a consequence of Lemma 1 and Lemma 6. Hence the proof is omitted.

\textbf{Lemma 7}. $\sum_k d_k^* \leq (1 + \frac{2 + \alpha}{\sqrt{10 + p}}) \cdot (1 + \pi) d^*$.

~

Using Lemmas 2, 3, 4, 6 and 7, we prove the following result, using a similar argument as for Theorem 5, but with different numbers.

\textbf{Theorem 8}. MP-PPTWC can be solved with an approximation ratio of $16 \ln 2 \cdot (1 + \epsilon + \frac{(2 + \alpha) \epsilon}{\sqrt{10 + p}}) (1 + \pi) \log T \simeq 46 (1 + \epsilon + \frac{(2 + \alpha) \epsilon}{\sqrt{10 + p}}) \log T$ in $O(n^{10 + p})$ time using Algorithm 2.

\textbf{Proof}. 
By Lemma 6, Algorithm 2 uses at most $m \leq (1 + \frac{2 + \alpha}{\sqrt{10 + p}}) m^*$ vehicles. 
By lemma 2, we obtain a reward of $\frac{1}{1 + \pi}$ of the optimal due to traveling distance approximation. 
By Lemmas 3 and 4, for each set, we obtain at least $\frac{1}{8 \ln 2 \log T (1 + \epsilon)}$ of the optimal reward, using the approximate algorithm. 
Hence, we get a revenue of at least 
\begin{equation}
\frac{1}{1 + \pi} \cdot \frac{\sum_{k=1}^{m} P_k^*}{(1 + (1 + \frac{2 + \alpha}{\sqrt{10 + p}}) \epsilon) 8 \ln 2 \log T} \geq \frac{1}{2(1 + \pi)} \cdot \frac{P^*}{8 \ln 2 (1 + \epsilon + \epsilon \frac{2 + \alpha}{\sqrt{10 + p}}) \log T}, 
\end{equation}
where $P^*$ is the optimal profit. 
Therefore, the approximation ratio is 
\begin{equation}
16 \ln 2 (1 + \pi) (1 + \epsilon + \frac{2 + \alpha}{\sqrt{10 + p}} \epsilon) \log T.
\end{equation}
By Lemmas 3 and 6, the running time is $O(n^{10 + p})$. \qed

\subsection{Algorithm 3}
\label{Algorithm 3}

\textbf{0}. Let $m = \frac{\sum_{i = 1 \dots n} q_i}{Q}$ and $s > 0$ such that the well-separated pair decomposition (WSPD) of $S$ with parameter $s$ has $m$ subsets

\textbf{1}. Compute the WSPD of $S$ with parameter $s$ into $W = (A_1, B_1), \dots, (A_m, B_m)$

\textbf{2}. Sort $W$ by $|A_i| + |B_i|$

\textbf{3}. For $i = 1$ to $m$ do

~~~~ If $\sum_{j \in A_i} q_j > Q$ or $R_{k-2}$ misses deadline of any sites from $A_i$ then 

~~~~~~~~ Label $A_i$ as "large"

~~~~~~~~ Else label $A_i$ as "small"

~~~~ If $\sum_{j \in B_i} q_j > Q$  or $R_{k-1}$ misses deadline of any site from $B_i$ then

~~~~~~~~ Label $B_i$ as "large"

~~~~~~~~ Else label $B_i$ as "small"

\textbf{4}. Let $k = 0, C = \emptyset$. 

\textbf{5}. For $i = 1$ to $m$ do

~~~~ If $A_i$ is a "small" subset, then 

~~~~~~~~ assign $A_i$ to vehicle $k$ and use the bi-criteria algorithm in \cite{Bansal} on (the complete graph of) $A_i$ to find an approximate route $R_k$ for vehicle $k$

~~~~~~~~ increment $k$ and set $C = C \cup A_i$

~~~~ If $C = S$ then stop

~~~~ If $B_i$ is a "small" subset, then 

~~~~~~~~ assign $B_i$ to vehicle $k$ and use the bi-criteria algorithm in \cite{Bansal} on (the complete graph of) $B_i$ to find an approximate route $R_k$ for vehicle $k$

~~~~~~~~ increment $k$ and set $C = C \cup B_i$

~~~~ If $C = S$ then stop

\textbf{6}. Return $k^* = k$ and the routes $R_{1 \dots k^*}$.

~

It is easy to verify that any route $R_k$ computed by Algorithm 3 is based on a subset of sites $A_i \in W$ or $B_i \in W$ , for some $i = 1 \dots m$. Thus, all routes comprise a WSPD of the original set of sites. Moreover, $R_k$ does not violate any capacity constraints. Also, note that there are only $k^* \leq 2 m = \frac{\sum_{i = 1 \dots n} q_i}{Q}$ vehicles used. This number of vehicles is always sufficient, regardless of the packing algorithm \cite{Dosa}.

We now analyze the performance of the algorithm. Let $m^*$ the optimal number of vehicles needed. Recall that $d_k^*$ is the optimal length of any tour that visits all sites in $S_k$ starting and ending at $D$, and $d^*$ the optimal total length of any set of $m^*$ such tours (over all possible $S_k$'s).

~

\textbf{Lemma 9}. $\sum_{k=1}^{2 m} d_k^* \leq \frac{2 m}{m^*} \frac{1 + \pi}{1 + s} d^*$.

\textbf{Proof}. 
Suppose sites are vertices of a regular polygon centered at the depot, as in the proof of Lemma 1. 
Recall that we denoted by $r$ the radius of the circumcircle of the sites, and by $l$ the side of the polygon. 
By the WSPD property of the routing, the distance between any two sites of different subsets is at least $s$ times the distance between any two sites from one of the subsets. 
We thus have $d^* \geq (n - m^*) l s + 2 m^* r$. 
Since $l = 2 r \sin(\pi/n)$, we have 
\begin{equation}
d^* \geq 2 m^* r (1 + s \sin(\pi/n) (n / m^* - 1)).
\end{equation}
On the other hand, 
\begin{equation}
\sum_{k=1}^{2 m} d_k^* \leq  ((n - 2 m) l + 2 r s) (2 m) \leq 4 m r (1 - 2 m \sin(\pi/n) + n \sin(\pi/n)).
\end{equation} 
Hence \begin{equation}
\sum_{k=1}^{2 m} d_k^* \leq \frac{2 m}{m^*} \frac{s - 2 m \sin(\pi/n) + n \sin(\pi/n)}{1 + s \sin(\pi/n) (n / m^* - 1)}.
\end{equation} 
Since $n / m^* >> 1$, we have 
\begin{equation}
1 + s \sin(\pi/n) (n / m^* - 1) > 1 + s.
\end{equation}
Also, 
\begin{equation}
1 - 2 m \sin(\pi/n) + n \sin(\pi/n) < 1 + s + \pi.
\end{equation}
Therefore, 
\begin{equation}
\sum_{k=1}^{2 m} d_k^* \leq \frac{2 m}{m^*} (1 + \frac{\pi}{1 + s}) d^*. 
\end{equation} \qed

~

\textbf{Lemma 10}. Using Algorithm 3, we get $\sum_{k=1}^{k^*} {d_k}^* \leq 2 (1 + \frac{\pi}{1 + s}) d^*$. 

\textbf{Proof}. 
It suffices to prove that $k^* \leq m^*$. Then, Lemma 10 follows directly from Lemma 9. 
Suppose, on the contrary, that $k^* > m^*$. In other words, there is a packing of the quantities supplied at the sites into $m^*$ bins of size $Q$. 
However, because of Step 4 of the algorithm, a WSPD with $2 m^*$ subsets would not cover the set of sites $S$ and it would need more than $2 m^*$ subsets to cover $S$. 
That is, the bin packing computed via WSPD has an approximation factor greater than 2, contradicting the result in \cite{Dosa}. \qed

~

We are now ready to prove the following result.

\textbf{Theorem 11}. Algorithm 3 solves MP-PPTWC within an approximation ratio of $11 (1 + \frac{\pi}{1 + s}) (1 + 2 \epsilon (1 + \frac{1}{1 + s})) \log T$. It runs in $O(s^2 n^{9 + p})$ time.

\textbf{Proof}. 
Note that WSPD creates a sub-network of $O(s^2 n_i)$ edges for each subset of size $n_i$.
Since Step 5 of the algorithm runs in $O(n_i^{8 + p} E)$ on a given graph $G$ of $n_i$ nodes and $E$ edges \cite{Bansal}, where $E$ is the number of edges in $G$, the running time is $O(s^2 n_i^{9 + p})$ per subset.
Since there are $k^* = O(n)$ subsets, and $\sum_{i=1}^{k^*} n_i = n$, we get a total running time of $O(s^2 \sum_{i = 1}{k^*} n_i^{9 + p}) = O(s^2 (\sum_{i = 1}{k^*} n_i)^{9 + p}) = O(s^2 n^{9 + p})$.
Using a similar argument as in Theorem 5, we get an approximation ratio of 
\begin{equation}
(1 + \frac{\pi}{1 + s}) + \epsilon \cdot \frac{\sum_{k=1}^m d_k^*}{d^*} 2 (1 + \frac{\pi}{1 + s}) \cdot \rho_{TW}, 
\end{equation}
where $\rho_{TW}$ is the factor given by the approximation ratio of the algorithm in Step 5. 
By Lemma 3, that would be $8 (\ln 2) \log T (1 + \epsilon)$. 
However, since the input graph of Step 5 is an $s-spanner$, we get an additional travel cost factor of $1 + 1/(1 + s)$
Hence, \begin{equation}
\rho_{TW} = 8 (\ln 2) \log T (1 + \epsilon (1 + 1/(1 + s))) \simeq 5.5 \log T (1 + \epsilon (1 + 1/(1 + s))). 
\end{equation}
Therefore, the overall approximation ratio is 
\begin{equation}
11 (1 + \frac{\pi}{1 + s}) (1 + 2 \epsilon (1 + \frac{1}{1 + s})) \log T.
\end{equation} \qed

~

\textbf{Note}. For sufficiently large $n$, since $s^2 = O(m) = O(\min{n, n^p})$, we get an approximation ratio of $11 (1 + 2 \epsilon) \log T$, for a running time of $O(\min{n^{10 + p}, n^{9 + 2 p}})$.

\section{Implementation and Experiments}
\label{Implementation}

\begin{figure}
	\includegraphics[scale=0.4]{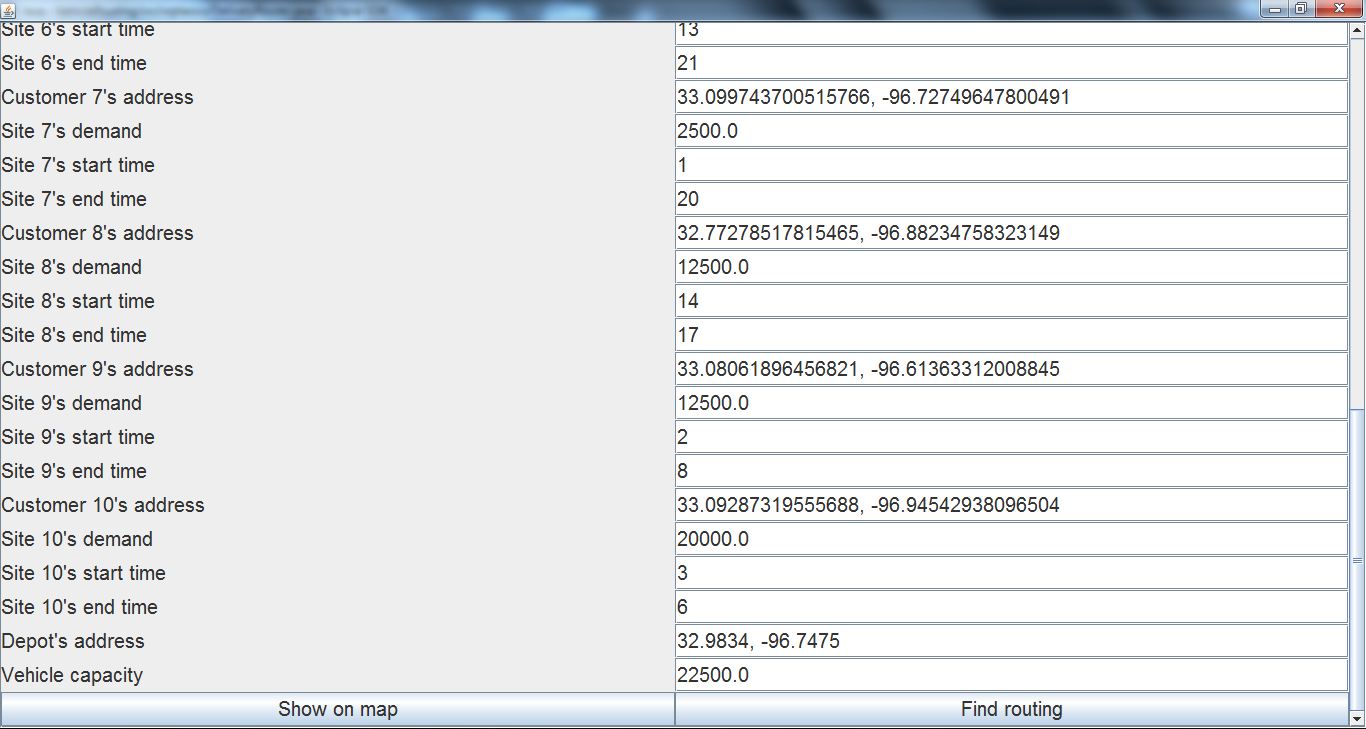}
	\caption{The GUI. Note that for all parameters of the problem, the GUI has a line with a label and a text field. }
\end{figure}

\begin{figure}
	\centering
	\includegraphics[width=0.6\textwidth, height=0.4\textwidth]{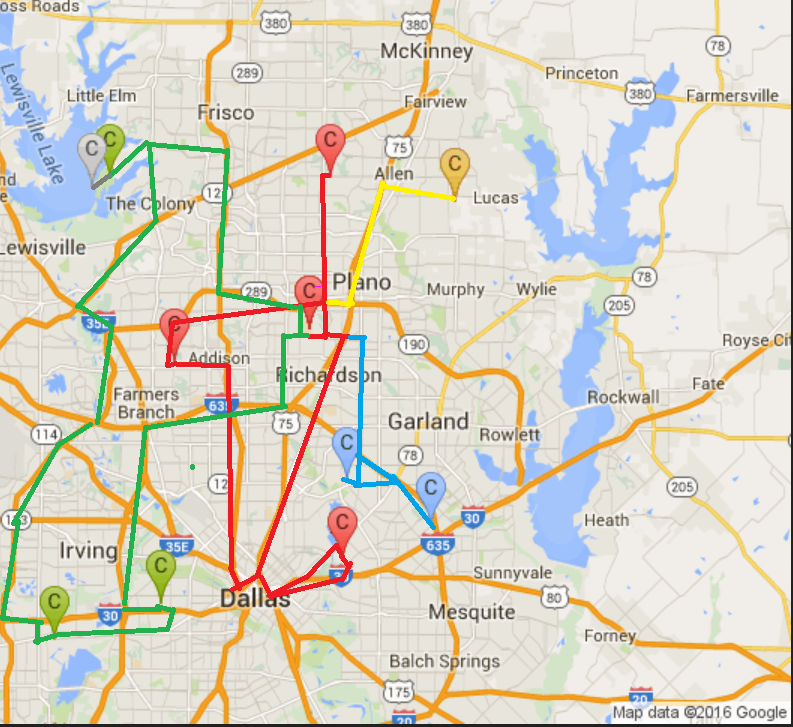}
	\caption{Routing for an instance where $n = 10$ is displayed with poly-lines of different colors on a Google Map. The depot is marked with red and supplier sites are marked with the colors of the routing visiting them.}
\end{figure}

We have implemented our first algorithm in JAVA, with Google Maps API, and simulated it on Google Maps.

The customers' locations are given as addresses. To find the point-to-point distances $d_{ij}$, we use Google Maps API to perform a Google Directions query from the address of customer $i$ to the address of customer $j$. The travel times $t_{ij}$ are also given by the Google Direction query. The Google Directions are given in JSON format. The routing is displayed on the Google Maps by constructing, for each vehicle, a multi-leg Google Directions query (i.e. a sequence of standard Google Direction queries with matching endpoints, except the start and the end points). In the resulting Google Map routing, only the paths are displayed on the map, the directions are not written.

Figure 2 shows the Graphical User Interface.

Figure 3 shows the how the routing computed by our algorithm is displayed on Google Maps. 


~

We have tested our algorithms on a dataset of randomly generated problem instances.

Tables 2 and 3 show, for each problem instance, the number of sites $n$, the maximum time window $T$, the running time $time$, and the performance ratio $\rho$ of Algorithm 1 (resp., Algorithm 3, for $s = 5$). 
The values $e_i$ and $l_i$ for each site were sampled uniformly at random in the intervals $[0, T/2]$, resp. $[T/2, T]$.
Here the peformance ratio is defined as $\rho = U/P$, where $P$ and $U$ are, respectively, the actual profit and the upper bound profit of the routing computed by the algorithm. 
These running times were obtained when running our program on a laptop with an Intel\textregistered Core I7-5500U processor at 2.40 GHz with 8 GB of RAM. 

Note that, in most cases, the approximation ratio is almost always below 2.5 for Algorithm 2 and below 2 for Algorithm 3, which is much smaller than the respective upper bounds for both algorithms. This reflects the average case scenario. 
In fact, for a given instance, the actual optimum may be lower than the upper bound, so the approximation ratio is even smaller for that instance. 
Moreover, the running time is also relatively fast in the average case, even for instances with 50 sites (rarely the running time exceeds a few minutes on these cases).
Thus, our algorithm is actually very practical. 

Also, note that the performance of Algorithm 1 is only slightly lower than that of Algorithm 3, in most instances. This is mainly due to the fact that the total traveling cost is, on average, much lower than the theoretical upper bound, especially for non-negligible $\epsilon$ values (i.e. ratio between costof a route and its reward).

Since these are the first algorithms for this particular problem, no comparison with other algorithms is needed.


~

~

\textbf{Table 2}. The performance ratio and running time of Algorithm 1, when run on different problem instances.

\begin{tabular}{|*{7}{l}|}
\hline problem & 			$n$ & 	$P$ &		$U$ &		$\rho$ &	$time$ (ms) &	$T$	\\
\hline Dallas\_wood\_10 &		10 &	66 &		117.5 &		1.78 &		299 &			15	\\
\hline Austin\_tools\_10 &		10 &	1497 &		2750 &		1.84 &		246 &			15	\\
\hline Denver\_stone\_10 &	10 &	153.6 &		307.2 &		2.0 &		41 &			12	\\
\hline Phoenix\_glass\_10 &	10 &	1662 &		3520 &		2.12 &		150 &			9	\\
\hline Dallas\_wood\_30 &		30 &	202.3 &		463 &		2.29 &		4766 &			15	\\
\hline Austin\_tools\_30 &		30 &	3338 &		5066 &		1.52 &		43006 &			15	\\
\hline Denver\_stone\_30 &	30 &	369 &		860 &		2.33 &		4422 &			12	\\
\hline Phoenix\_glass\_30 &	30 &	4518 &		8140 &		1.80 &		44498 &			9	\\
\hline Dallas\_wood\_50 &		50 &	282 &		570 &		2.02 &		278280 &		15	\\
\hline Austin\_tools\_50 &		50 &	4641 &		11747 &		2.53 &		20506 &			15	\\
\hline Denver\_stone\_50 &	50 &	551 &		1360 &		2.46 &		13392 &			12	\\
\hline Phoenix\_glass\_50 &	50 &	6749 &		16896 &		2.5 &		61992 &			9	\\
\hline
\end{tabular}

~

\textbf{Table 3}. The performance ratio and running time of Algorithm 3, when run on different problem instances.

\begin{tabular}{|*{7}{l}|}
\hline problem & 			$n$ & 	$P$ &		$U$ &		$\rho$ &	$time$ (ms) &	$T$	\\
\hline Dallas\_wood\_10 &		10 &	66.9 &		117.5 &		1.77 &		167 &			15	\\
\hline Austin\_tools\_10 &		10 &	1540 &		2750 &		1.79 &		135 &			15	\\
\hline Denver\_stone\_10 &	10 &	144 &		307.2 &		2.13 &		65 &			12	\\
\hline Phoenix\_glass\_10 &	10 &	1640 &		3520 &		2.14 &		57 &			9	\\
\hline Dallas\_wood\_30 &		30 &	297 &		463 &		1.55 &		1584 &			15	\\
\hline Austin\_tools\_30 &		30 &	3640 &		5066 &		1.39 &		47890 &			15	\\
\hline Denver\_stone\_30 &	30 &	555 &		860 &		1.55 &		1871 &			12	\\
\hline Phoenix\_glass\_30 &	30 &	6250 &		8140 &		1.30 &		30480 &			9	\\
\hline Dallas\_wood\_50 &		50 &	491 &		570 &		1.16 &		38153 &			15	\\
\hline Austin\_tools\_50 &		50 &	9390 &		11747 &		1.25 &		45260 &			15	\\
\hline Denver\_stone\_50 &	50 &	805 &		1360 &		1.69 &		16376 &			12	\\
\hline Phoenix\_glass\_50 &	50 &	13500 &		16896 &		1.25 &		125458 &		9	\\
\hline
\end{tabular}

\section{Conclusions and Future Work}
\label{Conclusions and Future Work}

We designed three approximation algorithms for the MP-PPTWC problem. We also implemented our algorithms and simulated the output of one of them using Google Maps. 

It remains as a future research problem to find approximation algorithms for more general versions of MP-PPTWC, which we describe now.

~

\textbf{One Vehicle per Supplier, Variable Supply (MP-PPTWC-VS)}. A supplier may be visited by at most one vehicle. The supply is variable and increases piece-wise linearly in time (under constant production rate $\rho_i > 0$). See figure 4 (right).

\begin{figure}
	\includegraphics[scale=0.5]{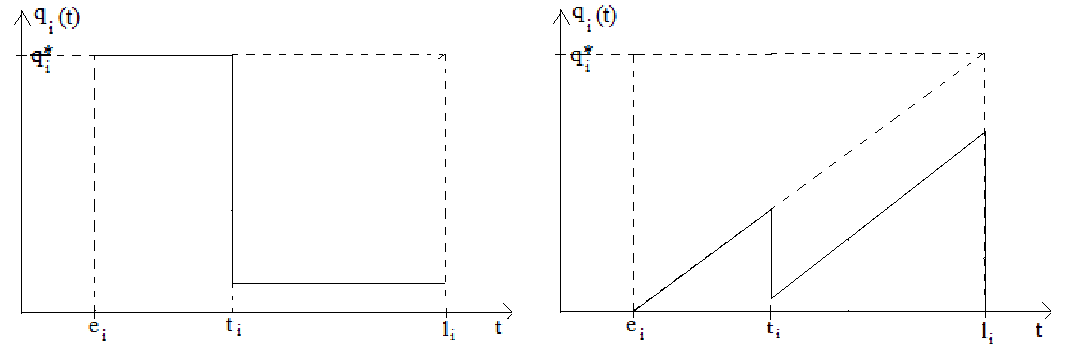}
	\caption{
Left: Standard version: One vehicle per supplier $i$, fixed supply. Some vehicle visits the supplier at time $t_i$ and picks up part of quantity $q_i$. 
Right: Version VS: One vehicle per supplier $i$, supply varying in time. Some vehicle visits the supplier at time $t_i$ and picks up part of quantity $q_i$}
\end{figure}

~

\textbf{Many Vehicles per Supplier, Fixed Supply (MP-PPTWC-MFS)}. A supplier may be visited by more than one vehicle. The supply is fixed (production rate $\rho_i = 0$) and thus piece-wise constant in time. See figure 5 (left).

~

\textbf{Many Vehicles per Supplier, Variable Supply (MP-PPTWC-MVS)}. A supplier may be visited by more than one vehicle. The supply varies in time and is piece-wise linearly increasing (constant production rate $\rho_i > 0$). See figure 5 (right).

\begin{figure}
	\includegraphics[scale=0.5]{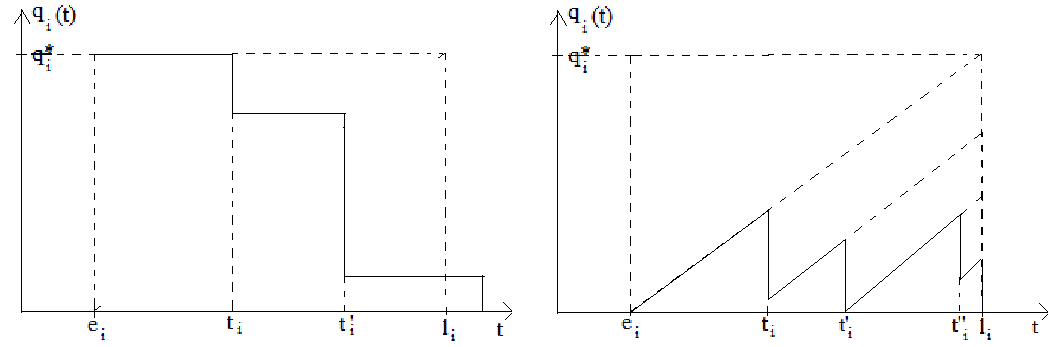}
	\caption{
Left: Version MFS: Many vehicles per supplier $i$, fixed supply. Some vehicles visit the supplier at times $t_i, t_i'$ and each picks up a part of the available quantity. 
Right: Version MVS: Many vehicles per supplier $i$, time-varying supply. Some vehicles visit the supplier at times $t_i, t_i', t_i''$ and each picks up a part (or all) of the available quantity}
\end{figure}

~

Note that our algorithms cannot be applied to these problems. First, the bin-packing step assumes that the supplies are constant, which is not the case for the VS version. Second, the reward-maximization approach for each vehicle assumes that a site can be visited by one and only one vehicle, so it does not work for the MFS, MVS versions.

\bigskip
\end{document}